\documentclass[preprint,showpacs,preprintnumbers,amsmath,amssymb]{revtex4}

\usepackage{graphicx}% Include figure files
\usepackage{dcolumn}% Align table columns on decimal point
\usepackage{bm}% bold math

\newcommand{\av}[1]{\left< #1 \right>}
\newcommand{\tens}[1]{\mathbf{#1}}
\newcommand{\vect}[1]{\mathbf{#1}}
\newcommand{\trace}{\mathrm{Tr} \,}
\newcommand{\ud}{\mathrm{d}}

\begin{document}

%\preprint{APS/123-QED}

\title{Coarse-grained simulations of flow-induced nucleation in
  semi-crystalline polymers}

\author{Richard S.~Graham$^1$ and Peter D.~Olmsted$^2$}%
 %\email{Richard.Graham@nottingham.ac.uk}
\affiliation{$^1$School of Mathematical Sciences, University of
  Nottingham, Nottingham NG7 2RD, UK.\\
$^2$School of Physics and Astronomy, University of Leeds,
  Leeds LS2 9JT, UK.
}%

\date{\today}

\begin{abstract}

We perform  kinetic Monte Carlo simulations of flow-induced nucleation in polymer melts with
an  algorithm that is tractable
even at low undercooling. The configuration of the non-crystallized
chains under flow is computed with a recent non-linear tube model. Our simulations predict
both enhanced nucleation and the growth of shish-like elongated nuclei for
sufficiently fast flows. The simulations predict several experimental phenomena and theoretically justify a previously empirical result for the flow-enhanced nucleation rate.
 The simulations are
highly pertinent  to both the fundamental understanding and process
modeling of flow-induced crystallization in polymer melts.
\end{abstract}


\maketitle

\section{ Introduction} The nucleation of microscopic crystallites in polymer liquids is
profoundly influenced by  flow \cite{KellerKolnaar97shrt,Binsgergen66}.
This flow-induced
crystallization (FIC),  is  a fascinating example of an externally driven,
non-equilibrium phase transition,  controlled by
kinetics.  Furthermore, FIC is ubiquitous in industrial
processing of semi-crystalline polymers,  the largest group of 
commercially useful polymers. A fundamental understanding of FIC
promises extensive control of polymer solid state properties, as virtually every property of  
practical interest is determined by the crystal morphology. Flow can drastically enhance
nucleation and trigger the formation of highly aligned,
elongated crystals, known as shish kebabs
\cite{KellerKolnaar97shrt}. Recent experiments on entangled
polymers, have studied, in detail, shish kebab formation
\cite{KimataSNKYKSK07shrt,HsiaoYSAZ05shrt,Balzano:2008p1800shrt} and the 
role of blend concentration \cite{SekiTOK02shrt}, molecular architecture
\cite{Heeley06shrt} and molecular relaxation time
\cite{Mykhaylyk:2008p1731shrt}. Often the most pronounced
flow-induced effects occur near the melting point, where quiescent
crystallization is  immeasurably slow \cite{Binsgergen66,Mykhaylyk:2008p1731shrt}.

The widely postulated mechanism for FIC states that flow forces the polymer chains
into elongated configurations, which lowers
the entropic penalty for crystallization \cite{KellerKolnaar97shrt}. However, this hypothesis has yet to be developed into a quantitative
molecular model.
% for both enhanced nucleation and shish
%formation.
FIC is extremely sensitive to the flow-induced  
configurations of the non-crystalline chains, so an accurate molecular flow
model is an essential prerequisite. Unfortunately,
most polymer flow models predict only the macroscopic stress tensor
and not the full molecular configuration. 
Alternatively,  detailed simulations of
polymer crystallization have provided much useful information on the
growth process \cite{Waheed:2005p1209,Hu:2002p1025,Zhang:2007p949}, yet
simulating primary nucleation has proven
difficult, especially at low undercooling, because of the extremely
long nucleation times.
At a much higher level of coarse-graining, models based on differential equations either assume an empirical dependence of the
nucleation rate on  the flow conditions
\cite{Eder97shrt}, the stress tensor
\cite{Zuidema:2001p1196}, or the chain stretch \cite{steenbakkers08:rheolflow}; or assert that free
energy changes 
under flow can be directly subtracted from the nucleation
barrier \cite{Coppola:2001p1012,Kulkarni:1999p2721}.
In either case the postulated FIC mechanism remains untested. An
intermediate level of coarse-graining is required to surmount these
difficulties.  

This letter presents coarse-grained kinetic Monte Carlo (MC) \cite{GILLESPIE:1977p3015}
simulations of anisotropic nucleation in flowing polymers. We compute chain configurations using a recent
molecular flow model \cite{graham03shrt} that
reliably predicts both neutron scattering \cite{bent03shrt,BlanchardPRLshrt} and bulk
stresses.
Our simulations predict both enhanced nucleation and
elongated shish nuclei.
Kinetic MC has previously been used to model quiescent crystal
growth in dilute polymers \cite{Doye:1998p1027}. However, it is
particularly suited to nucleation and our algorithm is
tractable even at low undercooling, providing an efficient and highly flexible
framework to simulate general anisotropic nucleation under external fields.

\section{Model}
 We compute the transient chain configuration
of the uncrystallized chains under flow using the GLaMM 
model \cite{graham03shrt}, with finite chain extensibility included 
using Cohen's approximation \cite{cohen91}. The chains are
divided into $Z$ sub-chains, each corresponding to an
entanglement segment of
$N_e$ Kuhn steps of length $b$. We take $N_e=100$ throughout this Letter. 
One deterministic run of the model
provides the end-to-end vector $\tens{f}_i(t)=\av{\vect{r}_i\vect{r}_i}$,
where the ensemble average is for sub-chains of type $i$. The data  for
an entire transient flow are used later in
the nucleation simulations.  All flow timescales are in units
of the subchain Rouse
time  $\tau_e$ and $\vect{r}$ is normalized by $\sqrt{N_e}b$.

 Deformation of the amorphous chains
has two effects on the nucleation kinetics: stretching reduces the entropic penalty for
crystallization; and monomer alignment modifies the probability of
compatible alignment with the nucleus. 
The
change in elastic free energy $\Delta F^{el}$ for chains with
ensemble average constraints
$\tens{f}=\av{\vect{r}\vect{r}}$, but locally at equilibrium, can be calculated by statistical mechanics \cite{OlmstedM94}. Although an analytic calculation is not
possible for finitely extensible chains, steep free energy gradients  in highly stretched chains suppress fluctuations. Thus our numerical calculations for uniaxial deformations show that $\Delta
F^{el}(\av{\vect{r}\vect{r}})$ can be accurately approximated by an
expression that interpolates between  Gaussian elasticity \cite{OlmstedM94} for
small $\trace{\tens{f}}$  
and Cohen's \cite{cohen91} approximation  with
$\vect{r}^2=\trace{\tens{f}}$ at high stretching,
\begin{equation}
  \label{eq:41}
  \Delta F^{el}= \frac{1}{2}\trace\tens{f}  - \frac{1}{2}\trace\ln\tens{f} -
  N_e\ln\left(1-\frac{\trace\tens{f}}{N_e}\right) .
\end{equation}
Similarly, 
numerical calculation of the monomer orientation distribution $w(\theta)$ for chains
with a constraint $\tens{f}$ are well approximated by using $\vect{r}^2=\trace{\tens{f}}$ in the expression for $w(\theta)$ derived from a direct constraint on $\vect{r}$ \cite{Jarecki07shrt}.
\begin{equation*}
\begin{array}{l}
  w(\theta) =
  \frac{\mathcal{L}^{-1}[\sqrt{\trace{\tens{f}}}/N_e]}{4\pi\sinh(\mathcal{L}^{-1}[\sqrt{\trace{\tens{f}}}/N_e])}
\cosh\left(\mathcal{L}^{-1}\left[\frac{\sqrt{\trace{\tens{f}}}}{N_e}\right]\cos\theta\right),
\end{array}
\end{equation*}
where $\mathcal{L}^{-1}$ is the inverse Langevin function and $\theta$ is the
angle between the monomer and  the principle axis of $\tens{f}$.
Each subchain has an individual $\tens{f}_i$ so is treated   as a separate species with
concentration $\phi_i=1/Z$.

\begin{figure}[htb]
  \centering
  \includegraphics[width=7.0cm]{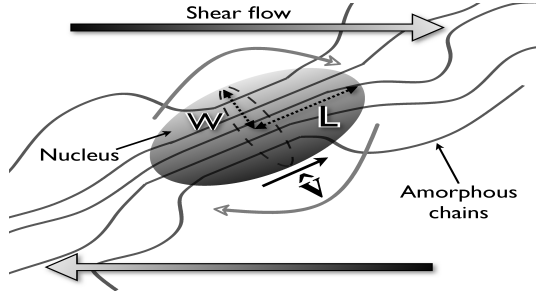}
  \caption{A sheared nucleus with protruding amorphous chains.}
  \label{fig:NucFig}
\end{figure}

Our coarse-grained simulations use the minimal nucleus
description required for anisotropic nucleation. The nucleus comprises $N_T$ crystallized
``monomers''  or Kuhn steps, 
 arranged in
stems, with each stem formed from a single chain (see fig~\ref{fig:NucFig}). The total
number of stems $N_s$ and the number  of monomers in each stem is
simulated. The arrangement of the 
monomers within the crystal is not resolved.  The nucleus is assumed to be
spheroidal with the polar radius $L$
parallel to the stems. 
Assigning a crystalline volume of $b_0^3$ to each monomer
and normalizing all lengths by $b_0$ gives the equatorial
radius $W=\sqrt{N_s/\pi}$, the volume $V=N_T$ and thus the polar
radius $L=\frac{3N_T}{4N_s}$. 
We also simulate the unit vector  $\hat{\vect{v}}$
parallel to the polar radius. As in classical nucleation theory the
nucleus free energy comprises the free energy of
crystallization,
proportional to the nucleus volume, and a free energy penalty
proportional to the surface area $S$. Thus the free energy in units
of $k_BT$ is $\mathcal{F}(N_T,N_s)=-\epsilon_BN_T+\mu_SS$, where
$\epsilon_B$ and $\mu_s$ are the coefficients of the volume and
surface area free energies, respectively.

\begin{figure*}[htb]
  \centering 
\includegraphics[width=16cm]{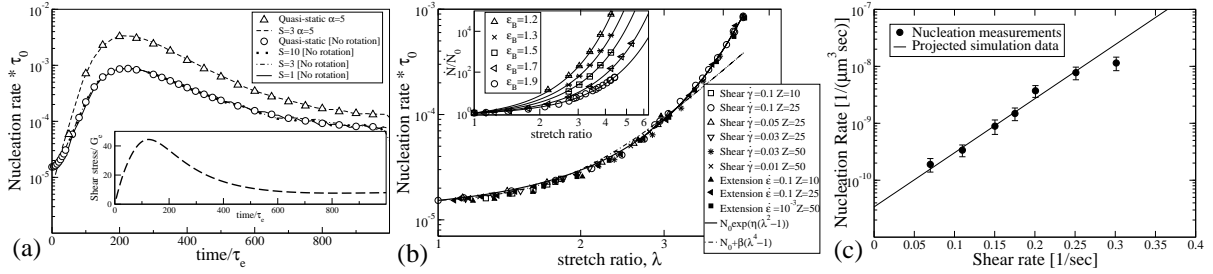}
  \caption{ (a) Transient and quasi-static nucleation rates for a $Z=25$ 
    monodisperse melt with
    $\epsilon_B=1.9$ and $\mu_S=1.9$ under start-up shear at
    $\dot{\gamma}\tau_e=0.1$; inset contains GLaMM model predictions for shear
    stress ($G_e$ is the shear modulus). (b)
    Master-curve of nucleation rate against stretch ratio 
    ($\dot{\epsilon}$ is the extension rate);
    inset shows the master-curves for varying $\epsilon_B$. (c) Steady-state nucleation rate measurements against shear rate for an industrial iPP melt \cite{Coccorullo:2008p2946} compared with projected simulation results.} 
\label{fig:NucData}
\end{figure*}

We simulate two types of MC moves: addition of a new stem containing one monomer and lengthening of an existing
stem by adding a new monomer, each with a corresponding
reverse move. As in \cite{Jarecki07shrt} we assume that to attach,
a monomer must be 
oriented within a solid tolerance angle $\Omega$ of the nucleus
orientation. 
%All monomers outside this angle are  unable to
%attach. 
For small $\Omega$,  the fraction of monomers within this angle is
$w(\theta)\Omega$, where $\theta$ is the angle between the nucleus
polar radius and the sub-chain principle strain axis. The stem
attachment rate for species $i$ is proportional to
its melt concentration $\phi_i$.
In contrast, for stem
lengthening only the next monomer along the
chain forming the stem can crystallize.  The  concentration of this
monomer at the nucleus surface
where the lengthening event occurs is taken to be unity. Thus the
attachment ($+$)
and detachment ($-$) move rates for stem addition ($k_{st}$) and stem
lengthening ($k_{len}$) are:
\begin{equation*}
  \begin{array}{rr}
    k^+_{st}=\frac{1}{\tau_0}\phi_i\omega_i\mathrm{min}(1,e^{-\Delta \mathcal{F}^+_{st}}) &
   k^-_{st}=\frac{1}{\tau_0}\mathrm{min}(1,e^{-\Delta \mathcal{F}^-_{st}})\\
k^+_{len}=\frac{1}{\tau_0}\omega_i\mathrm{min}(1,e^{-\Delta \mathcal{F}^+_{len}}) &
   k^-_{len}=\frac{1}{\tau_0}\mathrm{min}(1,e^{-\Delta \mathcal{F}^-_{len}})
\end{array}
\end{equation*}
where $\tau_0$ is the time for a monomer attachment attempt, 
$\omega_i=4\pi w_i(\theta_i)$ is the fraction of correctly
aligned monomers (normalized to unity in the quiescent limit), 
and the constant
$\ln(\Omega/4\pi)$ has been added to $\epsilon_B$.
The free energy change of attaching a new stem of species $i$ is
\begin{equation*}
\begin{array}{c}
\Delta
\mathcal{F}^+_{st}=\mathcal{F}(N_T+1,N_s+1)-\mathcal{F}(N_T,N_s)-\frac{1}{N_e}\Delta
F^{el}_i(\tens{f}_i),
\end{array}
\end{equation*}
where $\Delta F^{el}_i$ is the flow-induced free energy change in
sub-chain $i$. Similar calculations give the free 
energy changes for the other move types. 
A stem can only be detached if it contains a single monomer.

The kinetic MC algorithm requires a sum over all possible move
rates.  The area available for stem
addition moves is  $\mathcal{A}(N_s)$, which is taken to be proportional to
$\sqrt{N_s}$, 
and to give spherical nuclei in the
quiescent limit. To obey detailed balance the rate of stem removal
must be multiplied by $\mathcal{A}(N_s-1)/N_s$, the
probability of a given stem being at the nucleus surface. Each
stem can lengthen or shorten from either the top of bottom.
Thus the
total sum over all possible move rates is:
\begin{equation*}
  \label{eq:14}
  \begin{split}
  K_{\mathrm{Total}}&=\mathcal{A}(N_s)\sum_{i=1}^{Z}(k^+_{st})_i
+\frac{\mathcal{A}(N_s-1)}{N_s}\sum_{j=1}^{N_s}(k^-_{st})_{j}\\
&+\sum_{j=1}^{N_s}\Big(
   (k_{len}^{+top})_j + (k_{len}^{-top})_j
   +(k_{len}^{+bot})_j+(k_{len}^{-bot})_j\Big).
\end{split}
\end{equation*}
At each kinetic MC timestep one move is performed at random, with the
selection probability weighted by the move
rate. Time is then incremented by a
stochastically determined interval $\Delta t=-\ln\zeta/K_{Total}$,
where $\zeta$ is chosen uniformly on $[0,1]$ \cite{GILLESPIE:1977p3015,Doye:1998p1027}.

After each MC
step  the nucleus orientation
$\hat{\vect{v}}$ is incremented over  $\Delta t$ by
Brownian dynamics. Flow  
rotates the spheroid nucleus through the Jeffery algorithm
\cite{LEAL:1971p2453}. The angular diffusion time $\tau_{rot}$  scales with
both the nucleus volume $N_T$ and aspect ratio $\rho$
following the expressions in \cite{LEAL:1971p2453},  giving $\tau_{rot}=\alpha\tau_0
N_TG(\rho)$, where $G(\rho)$ depends only on the aspect ratio and we
have introduced a dimensionless constant $\alpha$,
connecting $\tau_{rot}$ with the
monomer attachment time $\tau_0$.
 As the Jeffery algorithm
is for Newtonian fluids we have
neglected non-Newtonian effects here. 
After each
time step all $\tens{f}_i$ values are updated from the
GLaMM model results, and $\Delta F^{el}_i$ and $\omega_i$ are
recalculated. The parameter $\mathcal{S}=\tau_e/\tau_0$
sets the ratio of flow and monomer attachment
timescales. We use $\mathcal{S}=10$ throughout this Letter, unless indicated
otherwise, although similar results
are obtained for any value of $\mathcal{S}>1$.

\section{Results} Each simulation evolves  a single nucleus
from a single monomer. The algorithm is
especially effective at low undercooling,
as for small nuclei the
rate sum is small, leading to large time steps.
The quiescent  free energy landscape can be calculated analytically from $\epsilon_B$ and $\mu_S$ to give a dimensionless nucleation
barrier $\Delta f^*$ and a critical nucleus of $n^*$ monomers. The simulated
nucleation time $\tau_N$ is the first time the polar and equatorial radii
simultaneously exceed the critical radius
$r^*=\sqrt[3]{3n^*/4\pi}$. Choosing a larger threshold size for
nucleation has little effect on $\av{\tau_N}$. 
 The results are accurately approximated by
$\av{\tau_N}=\tau_0\exp(\Delta f^*)$. We obtained good statistics for
barriers up to $25k_BT$ in $\sim50$hrs on one
$2$GHz processor, giving a nucleation time of $\sim10^{11}\tau_0$. The
nucleation times are Poisson distributed, so the nucleation rate can
be defined as
$\dot{N}_0=1/\av{\tau_N}$.

Under flow, the nucleation kinetics 
depend on the evolving chain deformation. We define the
instantaneous nucleation rate
 $ \dot{N}(t)\approx\frac{1}{1-n(t)}\frac{n(t+\Delta t) - n(t-\Delta
    t)}{2\Delta t}$
where $n(t)$ is the  cumulative fraction of runs nucleated  at
time $t$. 
For shear rates $\dot{\gamma}$ that are slow  compared to the critical nucleus rotation time 
($\tau_{rot}^{n^*}\dot{\gamma}\ll 1$) alignment effects can be
ignored,  equivalent to taking $\alpha=0$. 
The GLaMM model predictions  
for start-up of constant shear at $\dot{\gamma}\tau_e = 0.1$ of a $Z=25$
melt  and
the resulting instantaneous nucleation rate 
are in
fig~\ref{fig:NucData}(a). 
The
results are independent of the ratio of flow and crystallization
timescales for $\mathcal{S}\gtrsim 1$.
%hence the nucleation rate is independent of the earlier
%chain configurations.
In fact, fixing the non-crystalline chain
configuration to that corresponding to flow time $t$ for
the entire simulation and plotting the resulting quasi-static nucleation rate against $t$,
reproduces the transient results
(fig~\ref{fig:NucData}a). 
Thus nucleation  is fully controlled by the instantaneous
configuration of the surrounding chains, if
$\tau_e>\tau_0$. 

The
stretch ratio
$\lambda=\frac{1}{Z}\sum_{i=1}^{Z}\sqrt{\av{\vect{r}_i^2}}$ is the dominant factor
determining the  nucleation rate, despite variations in flow rate, molecular weight and
flow geometry (Fig~\ref{fig:NucData}(b)). 
This universal result is
somewhat surprising as the distribution of stretch
along the chain varies considerably with flow conditions, which may be expected to influence nucleation, especially if molecular weight and flow geometry is
varied. This result will be useful in deriving simple
differential models of FIC \cite{Zuidema:2001p1196,steenbakkers08:rheolflow} as 
the nucleation rate can be described the expression $\dot{N}=\dot{N}_0\exp(\eta(\lambda^2-1))$,
where $\eta$ is a
fitting constant. Also in fig~\ref{fig:NucData}(b) is the
curve $\dot{N}=\dot{N}_0+\beta(\lambda^4-1)$, which 
empirically fits measured nucleation rates from
flowing melts \cite{steenbakkers08:rheolflow}. The agreement with
our simulation data for $\lambda\lesssim 3.5$  theoretically
justifies this empirical expression. 
In the inset to fig~\ref{fig:NucData}(b) the
sensitivity of nucleation to chain stretching increases with
decreased undercooling (decreasing
$\epsilon_B$), as seen experimentally \cite{Coppola:2004p3046shrt}.
 Fig~\ref{fig:NucData}(c) directly compares  our
 simulations with steady-state nucleation rate measurements on a
 polydisperse isotactic polypropylene sample (iPP) during
 shear \cite{Coccorullo:2008p2946}. The parameter determination,
 projection of the simulation to large energy barriers and
 the approximation of the molecular weight distribution
 as a bimodal blend are detailed in appendix~\ref{sec:comp-with-nucl}. The close agreement shows our model can
 quantitatively account for FIC measurements.
 Fig~\ref{fig:NucData}(a)
 also shows the transient nucleation
rate when
$\alpha=5$.  This higher $\alpha$ value gives slower angular diffusion,  meaning that flow aligns sub-critical nuclei, 
accelerating nucleation. 
Here, the quasi-static nucleation remains
Poissonian and matches the transient
rates, although $\dot{N}$ now depends on both $\lambda$ and
$\dot{\gamma}\tau_{rot}^{n^*}$. Further increases of $\alpha$ give almost identical results.

\begin{figure}[htb]
\includegraphics[width=7.0cm]{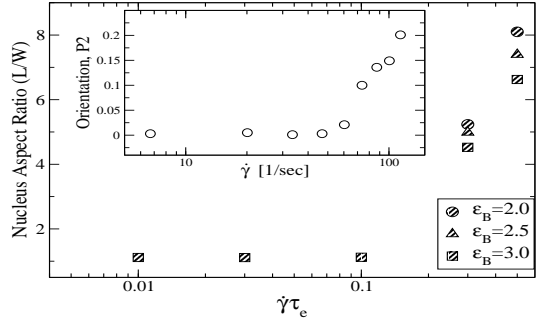}
  \caption{ Simulated nucleus aspect ratio at point of nucleation against shear rate for a $2\%$ high
    molecular weight blend, ($\alpha=5.0$, $\mu_S=2.5$, sheared for $120\tau_e$); inset shows experimental  orientation data on a bimodal blend sheared for $20$ sec \cite{Mykhaylyk:2008p1731shrt}.}
\label{fig:ShishData}
\end{figure}

In experiments shish nuclei are especially prevalent
in melts of short chains blended with a small amount
of very long chains
\cite{HsiaoYSAZ05shrt,Balzano:2008p1800shrt,SekiTOK02shrt,Heeley06shrt,
Mykhaylyk:2008p1731shrt}. We simulated a melt of $Z=15$ chains
blended with $2$wt$\%$ of  $Z=52$ chains, using
a generalization of the GLaMM model to
bimodal blends \cite{Graham:2009p3050shrt}. 
High shear produces very
elongated nuclei (fig~\ref{fig:ShishData}) from a purely
kinetic mechanism. The shish widen 
by adding new stems using any monomer from the melt, whereas  shish lengthen  
by adding monomers along an existing
stem. Therefore the concentration of monomers from stretched segments at the growth surface is
greater for lengthening than for widening, provided the nucleus
contains a disproportionate number of stretched segments. Fast flow conditions are
required for this disparity  to overcome the 
significant surface area cost of elongated
nuclei. Fig~\ref{fig:ShishData} (inset) shows experimental data for an
orientational order parameter, the P2 orientation function, against
shear rate for a $2\%$ high molecular weight blend
\cite{Mykhaylyk:2008p1731shrt}, highlighting similarities with our
predictions. 
Also in Fig~\ref{fig:ShishData} reduced undercooling, from reducing
$\epsilon_B$, increases the anisotropy,  as seen
experimentally \cite{BalzanoThesis08}.  When crystallization is less
favorable, the disparity in the kinetics of stretched and unstretched chain segments increases. 

\section{Discussion} Our efficient kinetic Monte Carlo algorithm for
flow-induced nucleation in polymer melts is tractable 
even at low undercooling.  The  flow-induced nucleation rate is a
universal function of the chain stretch ratio, independent of flow
rate, molecular weight and flow geometry, but with decreasing
undercooling causing increased flow sensitivity.  This universal
curve  is very similar to an empirical relationship that accurately
describes flow induced nucleation rate measurements
\cite{steenbakkers08:rheolflow}, justifying this empirical result. Our
successful quantitative comparison with nucleation measurements
\cite{Coccorullo:2008p2946} establishes that changes in chain free
energy  under flow can describe flow enhanced nucleation
in semi-crystalline polymers.  
In our bimodal blend simulations, a few percent of high molecular weight chains optimizes  shish formation and the degree of anisotropy increases with shear rate and decreased undercooling, all of which are seen experimentally \cite{HsiaoYSAZ05shrt,Balzano:2008p1800shrt,SekiTOK02shrt,Heeley06shrt,
Mykhaylyk:2008p1731shrt,BalzanoThesis08}.
The simulation can readily be generalized to
fully polydisperse melts, an essential step to  model industrial
polymer processing. The efficiency and
flexibility of our algorithm makes it suitable to simulate general anisotropic nucleation under external fields.

\section{Acknowledgements}
We thank the EPSRC for funding (GR/T11807/01) and L
Balzano, R Steenbakkers, G Peters, O Mykhaylyk, T Ryan and T McLeish for useful discussions.

\appendix

\section{Comparison with nucleation measurements.}
\label{sec:comp-with-nucl}

\subsection{Introduction}

Here we quantitatively compare our simulation results with directly
observed steady-state nucleation rates from a polymer melt under shear
\cite{Coccorullo:2008p2946}. These measurements, by Coccorullo {\em
  et. al}, were made on a commercial grade isotactic polypropylene
(iPP), known as iPP T30G.  These data show an exponential
dependence of the nucleation rate with shear rate, which our model correctly predicts. In order to quantitatively model this system, the required parameters
fall into two classes: molecular relaxation times, which can be
accurately calculated from literature values; and crystallization
parameters, for which order of magnitude estimates are provided by the
literature, but whose specific values we determine from the nucleation
data. The high polydispersity of iPP T30G is a further complication. As the
full molecular weight distribution (MWD) for this resin was not
reported, we choose a realistic MWD for highly polydisperse materials
and demonstrate that this distribution is consistent with both the
reported molecular weight averages and linear rheological
measurements. For non-linear flow modeling, we approximate this
polydisperse melt  as a bimodal blend, since no polydisperse
equivalent of the GLaMM mode is currently available. 
 We note that no fundamental modification of our simulation algorithm
 would be required to take advantage of the results of a polydisperse
 tube model. In this data comparison we are able to correctly describe these steady
state nucleation rate measurements against shear rate, using
crystallization parameters that are consistent with estimates provided
by literature measurements.

\subsection{Molecular relaxation times}

The  relaxation times of iPP T30G can be determined by the tube model
and confirmed against linear oscillatory shear measurements. Taking
the tube model parameters for iPP from the literature
\cite{vanMeerveld:2004p3039} and shifting $\tau_e$ to the experimental
temperature of $140^{\circ}$C using the time-temperature superposition
parameters for iPP T30G \cite{Coccorullo:2003p3040}, gives
$\tau_e=90$ns and $M_e=4.4$kg/mol. Here $\tau_e$ is the Rouse time of
an entanglement segment and $M_e$ is the molecular weight between
entanglements. The mass of a Kuhn step of iPP is $187.8$g/mol
\cite{Langston:2007p3049}, meaning the number of Kuhn steps per
entanglement segment is $N_e\approx25$. From these parameters the
chain Rouse time can be calculated using $\tau_R=(M/M_e)^2\tau_e$
\cite{doi86}. 

\begin{figure}[htb]
  \centering
 \includegraphics*[width=16cm]{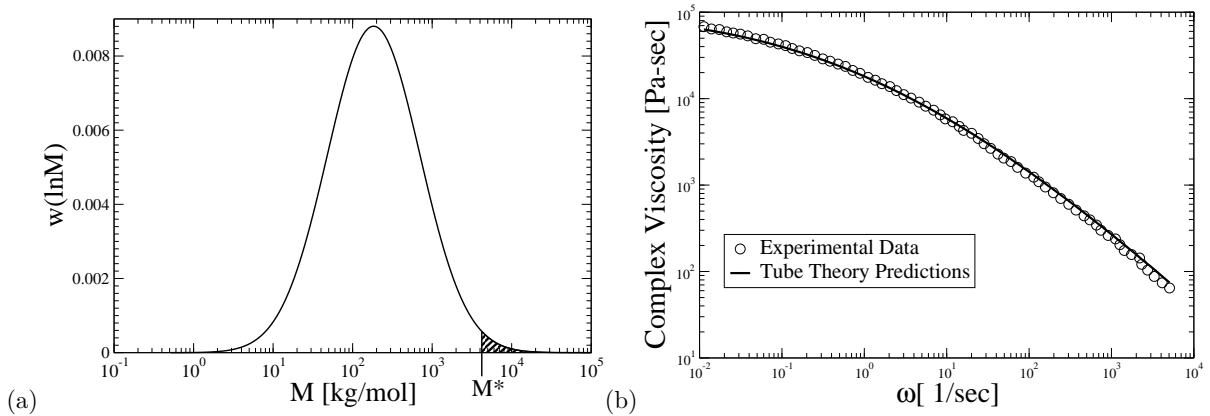}
  \caption{(a) Molecular weight distribution used to model iPP T30G calculated from eqn~(\ref{eq:4}). The shaded area  is the region of the molecular weight distribution defined as the high molecular weight tail in our modelling. (b) Linear rheological measurements on iPP T30G \cite{Coccorullo:2008p2946} along with the predictions of polydisperse tube model using the  molecular weight distribution discussed above and with $G_e=1.1\times10^6$Pa).}
  \label{fig:LinRheol}
\end{figure}

The form of the molecular weight distribution (MWD) for iPP T30G is
essential to our modeling as small amounts of high molecular weight
(HMW) material can control FIC (See, for example, ref
\cite{SekiTOK02shrt}). Unfortunately, the full MWD of iPP T30G was not
reported, with only the weight and number average molecular weights
being given \cite{Coccorullo:2008p2946}. Furthermore, standard
distribution functions, such as the log-normal and generalized
exponential distributions, typically under-estimate the HMW tail of
highly polydisperse melts \cite{VanRuymbeke:2002p1045}. Fortunately,
linear rheological measurements are highly sensitive to this HMW tail
at low frequencies and these data provide a clear upper-bound on the
amount of HMW material. Thus we select a realistic form for the full
MWD and show that this is consistent with both the reported complex
viscosity and molecular weight measurements ($M_w=481$kg/mol and
$M_n=75$kg/mol). We describe the MWD using a bimodal log-normal
distribution, since the augmented tail of this distribution accounts more accurately for
the HMW tail of highly polydisperse melts than standard monomodal
distributions. This takes the form 
\begin{equation}
  \label{eq:4}
  w_{Blend}(m)=(1-\phi_h)w_{LN}(m,M_{wl},M_{nl})+\phi_h w_{LN}(m,M_{wh},M_{nh})
\end{equation}
where $\phi_h$ is volume fraction of the HMW peak, $M_{wl/h}$ and
$M_{nl/h}$ are the weight and number average molecular weights for the
low and high molecular weight chains, respectively, and $w_{LN}$ is
the monomodal log-normal distribution. 
Taking $\phi_h=0.01\%$, $M_{wh}=3\times10^{5}$kg/mol,
$M_{nh}=2\times10^{4}$kg/mol, $M_{wl}=451$kg/mol and $M_{nl}=75$kg/mol
gives the correct overall weight and number average molecular weights
and correctly predicts the complex viscosity measurements. These
values were chosen as they correspond to the longest HMW tail that
correctly predicts the complex viscosity
measurements. Fig~\ref{fig:LinRheol}(a) shows the molecular weight
distribution and fig~\ref{fig:LinRheol}(b) compares the predicted
complex viscosity resulting from this distribution with measurements
on iPP T30G. Here the linear oscillatory shear response was computed
using a polydisperse tube model for linear response, which combines
the tube 
escape formula of Likhtman and McLeish \cite{likhtman02} with the 
Rubinstein and Colby algorithm for constraint release
\cite{rubinstein88} and is implemented within our in-house software
for molecular rheology, RepTate  
\footnote{
Ram\'irez, J., and A. E. Likhtman, Rheology of Entangled Polymers: Toolbox for the
Analysis of Theory and Experiments (Reptate, http://www.reptate.com,
2009).}. Using a monomodal log-normal distribution, with $\phi_l=0$,
produces slightly inferior agreement with the complex viscosity
measurements and cannot account for the nucleation measurements.

As discussed above, because there is no suitable polydisperse tube
model for non-linear flow, we approximate iPP T30G as a bimodal blend
when modelling its non-linear response. Fig~\ref{fig:LinRheol}(a)
shows that there is no clear molecular weight at which to define the
division between the two blend species. We choose the high molecular
weight fraction to be all material with a molecular weight above
$M^*=4200$kg/mol, giving a 
HMW volume fraction of $1\%$.  Taking a lower value of $M^*$ produces
a Rouse time that is too short to account for the flow nucleation
measurements, but we note that larger values of $M^*$ may be equally
valid choices. We note also that this fraction contains
discernible contributions from both of the monomodal terms in eqn~(\ref{eq:4}).
We compute the expectation value of the Rouse time of the high molecular weight tail via,
\begin{equation}
  \label{eq:1}
  \av{\tau_R}=\frac{\int_{M^*}^{\infty}\tau_e(m/M_e)^2w_{Blend}(m)\ud m}{\int_{M^*}^{\infty}w_{Blend}(m)\ud m},
\end{equation}
which gives $\tau_R=66$sec. All material below $M^*$ is defined as
matrix material, and a similar calculation gives the average Rouse
time of this matrix as
$2.1$ms. Thus, the matrix relaxes too quickly to be stretched by the flow
rates in the study of Coccorullo {\em et al.}. Hence, we model the non-linear flow dynamics of this
polydisperse material by approximating it as a bimodal blend
with the two fractions having the Rouse times discussed above. 

\subsection{Crystallization parameters}

To model iPP T30G we require its quiescent critical nucleus size and barrier height, at the experimental temperature.
An estimate for the number of Kuhn steps in a critical nucleus can be found from literature data. The critical nucleus diameter has been shown  to be of the order of the lamella thickness by transmission electron microscopy  \cite{bassett84,Olley:1989p3105} for iPP and  confirmed for a different polymer by atomic force 
microscopy \cite{Li:1999p3051}. For iPP the lamella thickness is $\sim
10$nm at $140^{\circ}$C. Using the iPP crystal density of
$0.9$g/cm$^3$ \cite{Ran:2001p3052} and the Kuhn step mass for iPP
\cite{Langston:2007p3049} and assuming a spherical critical nucleus
gives the number of Kuhn steps in a critical nucleus as $\sim 1000$. 
 
The barrier height can be estimated from the nucleation data of
Coccorullo {\em et. al} \cite{Coccorullo:2008p2946} (see
fig~\ref{fig:NucPlot}). Projecting these data to zero shear rate gives
a quiescent homogeneous nucleation rate of
$3.4\times10^{-11}/\mu$m$^3$/sec. This can be converted to a rate per
Kuhn step using the melt density of iPP  ($0.85$g/cm$^3$) and the iPP
Kuhn step mass. Thus  the density of Kuhn steps is
$\rho_K=2.7\times10^{9}/\mu$m$^3$ so the quiescent nucleation rate is
$\dot{N}_0=1.2\times10^{-20}$/sec per Kuhn step. Taking the
crystallization timescale of a Kuhn step $\tau_0$ to be of the order
of the Kuhn step relaxation time $\tau_K=\tau_e/N_e^2=0.144$ns, means
that   $\dot{N}_0\tau_0\sim10^{-30}$. Clearly nucleation in this
system is extremely rare and such a separation of timescales cannot
feasibly be simulated even by our highly efficient kMC
algorithm. Instead we systematically project our results to larger
energy barriers to allow a comparison with these data. 
\begin{figure}[htb]
  \centering
\includegraphics*[width=16cm]{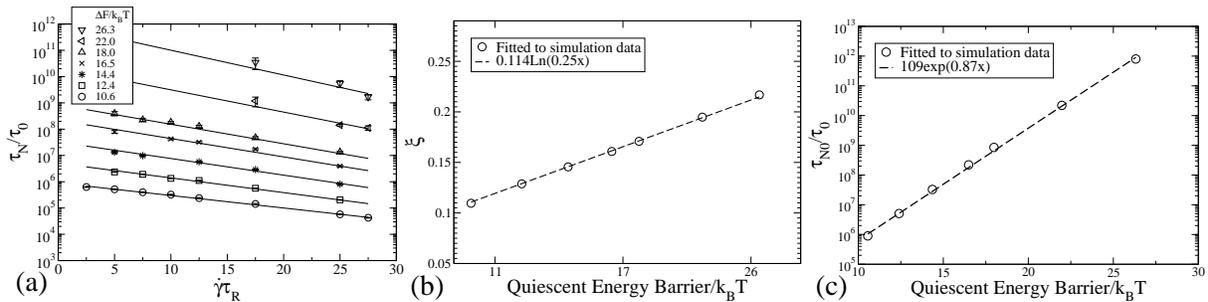}
  \caption{Projection of the simulated nucleation data to large energy barriers: (a) Simulation data for the average steady shear nucleation time against shear rate for a HMW volume fraction of $1\%$ fitted with eqn~(\ref{eq:3}) [shapes are simulation data and lines are fits]; (b) Variation of the sensitivity of the nucleation time to shear rate with quiescent barrier height; (c) Variation of $\tau_0$ with quiescent barrier height.}
  \label{fig:Pro}
\end{figure}

\subsection{Projection to large nucleation barriers}

To perform this projection we simulated the average nucleation time
over a range of quiescent nucleation barrier heights, varying from
$10.6$ to $26.3\mathrm{k_BT}$, with the critical nucleus size held
fixed at $1081$ Kuhn steps. Nucleus alignment was neglected for all of these simulations
($\alpha=0$, throughout). The steady-state configuration for both
blend components for various shear rates was calculated using a recent
generalisation of the GLaMM model to bimodal blends
\cite{Graham:2009p3050shrt}. The dimensionless steady-state nucleation
time $\tilde{\tau}_N=\tau_N/\tau_0$ against shear rate $\dot{\gamma}$
was simulated using the GLaMM model predictions, in steady state, for
a range of shear rates and molecular weights. The results are
independent of molecular weight when the shear rate is expressed as a
Rouse Weissenberg number $\dot{\gamma}\tau_R$, but do depend on the
height of the quiescent nucleation barrier, $\Delta f^*$, as shown in
figure~\ref{fig:Pro}(a). The simulated nucleation times vary
exponentially with shear rate and can be described by the expression 
\begin{equation} 
  \label{eq:3}
 \tilde{\tau}_N(\dot{\gamma})=\tilde{\tau}_{N0}\exp(-\xi\dot{\gamma}\tau_R),
\end{equation}
where $\tilde{\tau}_{N0}$ and $\xi$ are fitting parameters characterizing the decay of the nucleation time with shear rate, both of which depend on the barrier height $\Delta f^*$. 
The lines in fig~\ref{fig:Pro}(a) are the result of fitting
eqn~(\ref{eq:3}) to the simulation data. Both of the fitting
parameters have a simple dependence on $\Delta f^*$, as shown in
figs~\ref{fig:Pro}(b) and (c). Data for to the two largest quiescent
barriers was used only to confirm the projection and was not involved in
the fitting.  This procedure allows the simulation results to be projected to
nucleation barriers that are too high to be simulated.

\subsection{Data comparison}

\begin{figure}[htb]
  \centering
  \includegraphics*[width=7cm]{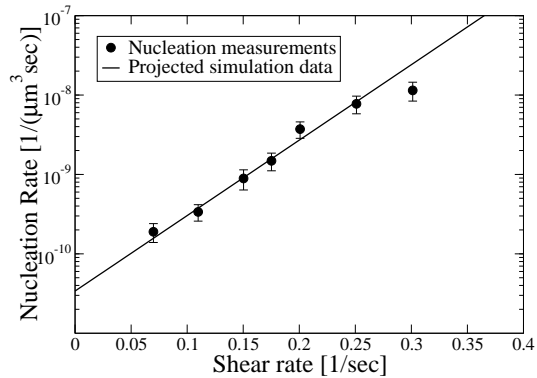}
  \caption{Comparison of measured steady shear nucleation rate against shear rate for an iPP resin \cite{Coccorullo:2008p2946} with our projected simulation results.}
  \label{fig:NucPlot}
\end{figure}

We can use our projected results to model the data of Coccorullo {\em
  et. al} \cite{Coccorullo:2008p2946}. The value of the projected zero
shear nucleation rate from these data, combined with the projection
formula in fig~\ref{fig:Pro}(c) and the requirement that
$\tau_0\sim\tau_K$, gives a quiescent nucleation barrier of $\Delta
f^*\sim70\mathrm{k_BT}$. The final value of $f^*$ is chosen
to reproduce the slope of the experimental data in
figure~\ref{fig:NucPlot}. Thus, using the Rouse time computed above
and taking a value of  $\Delta f^*=72\mathrm{k_BT}$ gives $\xi=0.33$
from the projection formula in figure~\ref{fig:Pro}(b). This
quiescent barrier specifies
$\tilde{\tau}_{N0}=1.74\times10^{29}$ by the projection formula in
figure~\ref{fig:Pro}(c). Thus taking $\tau_0=0.46$ns produces
agreement with the experimental data in
fig~\ref{fig:NucPlot} extrapolated to zero shear rate. We can also confirm that this
value for $\tau_0$ is
$\sim\tau_K$. With these parameters and using
$\dot{N}(\dot{\gamma})=1/\tau_N(\dot{\gamma})$, which is valid because
all of the simulated nucleation processes are Poissonian,
eqn~(\ref{eq:3}) can be written as 
\begin{equation}
  \label{eq:2}
  \dot{N}(\dot{\gamma})=\frac{\rho_K}{\tilde{\tau}_{N0}\tau_0}\exp(\xi\dot{\gamma}\tau_R)=3.4\times10^{-11}\exp(21.8\mathrm{sec}\dot{\gamma})/\mathrm{sec}/\mu\mathrm{m}^3.
\end{equation}
Eqn~(\ref{eq:2}) produces the  projected simulation data in
fig~\ref{fig:NucPlot}.

In summary, our model correctly predicts the exponential dependence
of nucleation rate on shear rate as observed by Coccorullo {\em et
  al.} \cite{Coccorullo:2008p2946}. Furthermore, with a characterization of the
full molecular weight distribution and extrapolation of our simulation
results to high nucleation barriers, we can make a
direct quantitative  comparison with these data. We use
crystallization parameters that are estimated from the literature but whose final values are
determined in response to the nucleation data. This results in very good
agreement with directly measured nucleation rates under flow using
parameters that are consistent with literature estimates.

Several consequences arise from this data comparison. It demonstrates that
changes in chain free energy upon stretching under flow are sufficient
to quantitatively account for the degree of enhanced nucleation
measured in flowing semi-crystalline polymers. It also indicates the
strong need for comprehensive material characterization to accompany
FIC measurements. In particular, an accurate characterization of the
sample's high molecular weight tail, along with linear rheological
modeling, will clearly define the full spectrum of molecular relaxation
times. The issues of polydispersity could be solved either through FIC
experiments on model polymers or through the emergence of a reliable
non-linear flow model for polydisperse polymers. In principle, our
model can predict the effect of varying molecular weight distribution,
temperature and flow history, including transient flows, and a more
comprehensive test of the model's predictions could be produced by
systematic comparison to such a data set. The close agreement between theory
and experimental data and our comparison provides a foundation on
which further quantitative comparisons can be built.

\end{document}